\documentclass[prd,preprint,showpacs,aps,epsfig]{revtex4}
\usepackage{latexsym}
\usepackage{color}
\usepackage{graphicx}
\usepackage{subfig}
\usepackage{amsmath,amssymb}
\usepackage{makeidx}
\usepackage{hyperref}
\usepackage{subeqnarray}
\usepackage[utf8]{inputenc}
\usepackage{indentfirst}
\usepackage{booktabs}
\usepackage{multirow}
\usepackage{array}
\usepackage{float}
\usepackage{ulem}
\usepackage[papersize={9in,13in}]{geometry}

\graphicspath{{Figuras/}}

\makeindex

\begin{document}

\title{Radiant gravitational collapse with anisotropy in pressures and bulk viscosity}
	
\author{A. C. Mesquita}
\email{arthurcamara2007@hotmail.com}
\author{M. F. A. da Silva}
\email{mfasnic@gmail.com}
	
\affiliation{Departamento de F\' {\i}sica Te\' orica, Universidade do Estado do Rio de Janeiro, Rua S\~ ao Francisco Xavier $524$, Maracan\~ a, CEP 20550--013, Rio de Janeiro -- RJ, Brazil}
	
\date{\today}
	
	
\begin{abstract}
	 We model a compact radiant star that undergoes gravitational collapse from a certain initial static configuration until it becomes a black hole. The star consists of a fluid with anisotropy in pressures, bulk viscosity, in addition to the radial heat flow. A solution of Einstein's field equations with temporal dependence was presented to study the dynamic evolution of physical quantities, such as the mass-energy function, the luminosity seen by an observer at infinity and the heat flow. We checked the acceptability conditions of the initial static configuration to obtain a range of mass-to-radius ratio in which the presented star model is physically reasonable. The energy conditions were analyzed for the dynamic case, in order to guarantee that the model is composed of a physically acceptable fluid within the range of the mass-to-radius ratio obtained for the static configuration or if they will be modified during the collapse.
\end{abstract}

	
\maketitle
	
\section{Introduction}
 
 Neutron stars can be detected through optical and X-ray observations, which reveal properties crucial for understanding their structure and evolution, such as surface radius and temperature \cite{LattimerB}. The internal structure of a neutron star depends on its equation of state, that is, a relationship between density and pressure within it. However, as we have not yet been able to produce such high densities in laboratories, we are not aware of the equation of state that best describes this matter, making its theoretical modeling difficult. Analogous to a white dwarf, this type of star has an upper limit of mass at which it would fall out of equilibrium and continue to collapse. Oppenheimer and Volkoff \cite{OppennheimerA}, using the theory of general relativity, established an upper limit of 0.7$M_\odot$, which became known as the Tolman-Oppenheimer-Volkoff limit, or simply TOV limit. Modern estimates shift this upper bound to about 2$M_\odot$ \cite{Cameron1959}. Soon after the TOV limit was established, Oppenheimer and Snyder \cite{OppennheimerB} studied the cataclysmic behavior for neutron stars with masses greater than this limit. The star will contract until its surface radius approaches $r$=$2m$ ($m$ is the mass of the central object), the Schwarzschild radius. When it exceeds this Schwarzschild radius, no information is transmitted to the region outside $r$. Thus, the fate of a neutron star whose mass exceeds the TOV limit is a black hole.

The observational scenario has shown to be very promising in bringing us new possibilities for the study of these compact objects. From the discovery of the spiralization and coalescence of a binary neutron star system (GW170817), by the Laser Interferometer Gravitational Waves Observatory (LIGO, VIRGO) (on August 17, 2017), a new alternative for accessing the equation of state emerged of such stars at high densities \cite{AbbottB, AbbottA}, at least to exclude some of them. This new astrophysical observation window, accessed through gravitational waves, has brought us many surprises and new challenges. In another work \cite{AbbottC}, it is observed what appears to be the coalescence of a binary system involving a black hole of about 22.2 - 24.3 $M_\odot$ and a compact object of approximately 2.50 - 2.67 $M_\odot$ , the latter having a mass too small to be a black hole, but larger than expected so far for a neutron star. 

Expanding our knowledge about the behavior of fluids under strong self-gravitation in light of the TGR is crucial for the interpretation of results like this, which emerge from the new observations.

In this article we are interested in investigating the temporal evolution, along the collapse process, of certain physical quantities, starting from an initial static configuration to the formation of a black hole, through the introduction of a temporal dependence in the metric. We consider a spherically symmetrical distribution of a fluid with anisotropy in pressures, with heat flux and viscosities, governed by a non-local equation of state, originally proposed by Hèrnandez and Núñez \cite{HernandezA}.

\ Our article is organized as follows. In section 2, we present a description of the geometry of spacetime and the energy-momentum tensor. In section 3, we present a special metric, time dependent, to study the evolution of an initial static configuration, for which the field equations lead to a non-local equation of state for fluids with anisotropic pressures. Then, in section 4, we investigate the evolution from collapse to black hole formation from a solution given by a density profile proposed by Wyman \cite{Wyman} that is written in a similar way to the one presented in Hernández and Núñez \cite{HernandezA}. We also corrected some results obtained from the latter, exploring the graphic behavior of quantities such as mass-energy enclosed in the surface of the distribution, luminosity for an observer at infinity, effective surface temperature, adiabatic index effective, heat flux and scalar expansion. In section 5, the energy conditions for the dynamic case are analyzed. Finally, in section 7 we present our final remarks. We include an Appendix presenting the energy conditions considered here.

\section{Einstein's Field Equation}

In order to study the gravitational collapse problem, we need to separate spacetime into three regions: the first consists of the interior region, that is, the spherically symmetric distribution of matter. The second, an outer region which is  fullfiled by null radiation, emitted by the matter distribution. Finally, the third of them refers to a $\Sigma$ junction hypersurface that separates these last two.

Let $g_{ij}$ be the metric intrinsic to the hypersurface $\Sigma$, which takes into account the description in comoving coordinates of the inner  spacetime, that is
\begin{equation}
    \label{1}
    ds^2_\Sigma=g_{ij}d\xi^i d\xi^j=-d\tau^2 + R^2(\tau)d\Omega^2 \ ,
\end{equation}
where $d\Omega^2= d\theta^2 +\text{sen}^2\theta d\phi^2$ is the angular element and $\tau$ represents the proper time, with $\xi^i= (\tau,\theta,\phi)$ representing the coordinates intrinsic to $\Sigma$.

On the other hand, the interior space-time of the matter distribution, is described by a spherically symmetric metric in the most general way possible, using comoving coordinates, given by

\begin{equation}
 \label{2}
    ds_-^2= g^-_{\alpha\beta}d\chi_-^\alpha d\chi_-^\beta =-A^2(r,t)dt^2 + B^2(r,t)dr^2+C^2(r,t)d\Omega^2 \ ,
\end{equation}
where $\chi_-^\alpha=(\chi_-^0,\chi_-^1,\chi_-^2,\chi_-^3)=(t,r,\theta,\phi)$ are the coordinates of the interior space-time.

The energy-momentum tensor, describing the matter that fills such space-time is represented by
\begin{eqnarray}
\label{5}
    T^-_{\alpha\beta} &=& (\rho+P_\perp)u_\alpha u_\beta +P_\perp g_{\alpha\beta}+(P_r-P_\perp)X_\alpha X_\beta + q_\alpha u_\beta \\ \nonumber &+&q_\beta u_\alpha  -2\eta\sigma_{\alpha\beta}-\zeta\Theta(g_{\alpha\beta}+u_\alpha u_\beta) \ ,
\end{eqnarray}
where $\rho$ is the energy density of the fluid, $P_r$ is the radial pressure, $P_\perp$ is the tangential pressure, $X_\alpha$ is a unit 4-vector along the radial direction, $u ^\alpha$ is the 4-velocity and $q_\alpha$ is the radial heat flux vector, which satisfy $q_\alpha u^\alpha=0$, $X_\alpha X^\alpha=1$, $X_\alpha u^\alpha=0$ and $u^\alpha u_\alpha=-1$. The 4-vectors are given by $u^\alpha=\delta^\alpha_0/A$,  $q^\alpha=q\delta_1^\alpha$ and $X^\alpha=\delta^\alpha_1/B$. The amounts $\eta>0$ and $\zeta>0$ are the shear viscosity and volume viscosity coefficients, respectively. Whereas $\sigma_{\alpha\beta}$ and $\Theta$ are, respectively, the shear tensor and the expansion scalar.

On the other hand, let us now consider the outer spacetime described by the Vaidya metric \cite{VaidyaB},  written as

\begin{equation}
 \label{3}
{ds_+}^2 = {g_{\alpha\beta}}^+d\chi_+^\alpha d\chi_+^\beta=-\left(1-\frac{2m(v)}{\textbf{r}}\right)dv^2-2dvd\textbf{r} +\textbf{r}^2d\Omega^2
\end{equation}
where $\chi_+^\alpha=(\chi_+^0,\chi_+^1,\chi_+^2,\chi_+^3)$ are the coordinates of outer spacetime and $m(v)$ represents the total fluid energy stored within the hypersurface $\Sigma$ as a function of the time delay $v$. 

The energy-momentum tensor for the outer region, representing a pure radiation field, is given by
\begin{equation}
 \label{4}
    T^+_{\alpha\beta}= ek_\alpha k_\beta \ ,
\end{equation}
where $k^\alpha$ is a null vector and $e$ is the radiation energy density measured locally by an observer over $\Sigma$. Hereafter we use the indices "+" or "-" to represent quantities referring to outer and inner spacetime, respectively. 

We can write the expansion scalar, the shear tensor, and the shear scalar, respectively, as
\begin{equation}
 \label{6}
    \Theta=u^\alpha_{;\alpha}=\frac{1}{A}\left(\frac{\dot{B}}{B}+\frac{2\dot{C}}{C}\right) \ ,
\end{equation}

\begin{equation}
 \label{7}
    \sigma_{\alpha\beta}=u_{(\alpha;\beta)}+\dot{u}_{(\alpha}u_{\beta)}-\frac{1}{3}\Theta(g_{\alpha\beta}+u_\alpha u_\beta) \ ,
\end{equation}

\begin{equation}
  \label{8}
    \sigma=-\frac{1}{3A}\left(\frac{\dot{B}}{B}-\frac{\dot{C}}{C}\right) \ ,
\end{equation}
where the parentheses in the subscripts in ($\ref{7}$) mean symmetrization, the dot in ($\ref{6}$) and ($\ref{8}$) represents $\partial/\partial t$ e $\dot{u}_\alpha=u_{\alpha;\beta}u^\beta$.

Combining the metric (\ref{2}) with the energy-momentum tensor (\ref{5}), the Einstein's field equations for the inner region are given by
\begin{eqnarray}\small
  \label{9}
    -\left(\frac{A}{B}\right)^2\left[2\frac{C''}{C}+\left(\frac{C'}{C}\right)^2-2\frac{C'}{C}\frac{B'}{B}\right]+\left(\frac{A}{C}\right)^2 + \frac{\dot{C}}{C}\left(\frac{\dot{C}}{C}+2\frac{\dot{B}}{B}\right)
    = kA^2\rho \ ,
\end{eqnarray}
\begin{eqnarray}
  \label{10}
   \frac{C'}{C}\left(\frac{C'}{C}+2\frac{A'}{A}\right)-\left(\frac{B}{C}\right)^2 -\left(\frac{B}{A}\right)^2\left[2\frac{\Ddot{C}}{C}+\left(\frac{\dot{C}}{C}\right)^2-2\frac{\dot{A}}{A}\frac{\dot{C}}{C}\right] \nonumber \\
    = kB^2(P_r+4\eta\sigma-\zeta\Theta) \ ,
\end{eqnarray}
\begin{eqnarray}
 \label{11} 
  \left(\frac{C}{B}\right)^2\left[\frac{A''}{A}+\frac{C''}{C}-\frac{A'}{A}\frac{B'}{B}+\frac{A'}{A}\frac{C'}{C}-\frac{B'}{B}\frac{C'}{C}\right] \nonumber \\
   + \left(\frac{C}{A}\right)^2\left[-\frac{\Ddot{B}}{B}-\frac{\Ddot{C}}{C}+\frac{\dot{A}}{A}\frac{\dot{B}}{B}+\frac{\dot{A}}{A}\frac{\dot{C}}{C}-\frac{\dot{B}}{B}\frac{\dot{C}}{C}\right] \nonumber \\
   =kC^2(P_\perp-2\eta\sigma-\zeta\Theta) \ ,
\end{eqnarray}
\begin{equation}
 \label{13}
   2\frac{A'}{A}\frac{\dot{C}}{C}+2\frac{\dot{B}}{B}\frac{C'}{C}-2\frac{\dot{C}'}{C} =- kAB^2q \ ,
\end{equation}
where $k=8\pi$ in the geometric coordinate system ($c=G=1$), the dot represents $\partial/\partial t$, while the line represents $\partial/\partial r$.

\section{Dynamic solution of field equations}

Just like in \cite{Veneroni} and \cite{Pretel}, we introduce a time dependent function in the static metric proposed by Hernández and Núñez \cite{HernandezA} in order to study the evolution of gravitational collapse from a given initial configuration, that is,
\begin{equation}
 \label{27}
    ds^2_-=-\frac{\xi^2}{h(r)}dt^2+\frac{f(t)}{h(r)}dr^2+f(t)r^2d\Omega^2 \ ,
\end{equation}
where $\xi$ is an arbitrary constant.

With the metric written in this way we can see that, according to (\ref{2}),
\begin{equation}
\label{28}
     A^2(r,t)=\frac{\xi^2}{h(r)} \ ,\  B^2(r,t)=\frac{f(t)}{h(r)} \ , \  C^2(r,t)=r^2f(t) \ .
\end{equation}

In this way, the expansion ($\ref{6}$) is given by 

\begin{equation}
\label{29}
\Theta=\frac{3h^{1/2}}{2\xi}\frac{\dot{f}}{f}\,
\end{equation}
and the shear scalar ($\ref{7}$) is zero, provided that $\dot{B}/B=\dot{C}/C$.

Thus, from the field equations (\ref{9})-(\ref{13}) we obtain the energy density, the radial and tangential pressures and the heat flux as follows:
\begin{equation}
 \label{30}
    8\pi\rho= \frac{1-h-rh'}{fr^2}+\frac{3h}{4\xi^2}\frac{\dot{f}^2}{f^2} \ ,
\end{equation}
\begin{equation}
 \label{31}    
    8\pi(P_r-\zeta\Theta)= \frac{h-rh'-1}{fr^2}+\frac{h}{4\xi^2}\frac{\dot{f}^2}{f^2}-\frac{h}{\xi^2}\frac{\Ddot{f}}{f} \ ,
\end{equation}
\begin{equation}
 \label{32}
    8\pi(P_\perp-\zeta\Theta)= \frac{h'^2-hh''}{2hf}+\frac{h}{4\xi^2}\frac{\dot{f}^2}{f^2}-\frac{h}{\xi^2}\frac{\Ddot{f}}{f} \ ,
\end{equation}
\begin{equation}
  \label{33}
    8\pi q=\frac{h'h^{1/2}}{2\xi}\frac{\dot{f}}{f^2} \ .
\end{equation}

If we consider $f(t)=1$ and $\dot{f}(t)=0$, we obtain the field equations for a static fluid, corresponding to the equations obtained by Hernández and Núñez \cite{HernandezA}. Thus, $f(t)=1$ and $\dot{f}(t)=0$ will represent the instant when the fluid starts to collapse, as will be seen in the next section.

Now, considering the equations ($\ref{27}$), we can rewrite the expression ($\ref{20}$) and  ($\ref{24}$), from the Appendix, as
\begin{equation}
  \label{34}
     m =\left[\frac{rf^{1/2}}{2}\left(1+\frac{hr^2\dot{f}^2}{4\xi^2 f}-h\right)\right]_\Sigma \ ,
\end{equation}
\begin{equation}
\label{35}
(qB)_\Sigma=(P_r-\zeta\Theta)_\Sigma \ .
\end{equation}

The luminosity seen by an observer at rest at infinity can be rewritten by replacing the functions ($\ref{28}$) and ($\ref{31}$) into ($\ref{26}$), as ,
\begin{equation}
 \label{36}    
    L_\infty=\left[\frac{h}{8\xi^2f}\left(h-rh'-1+\frac{r^2h\dot{f}^2}{4\xi^2f}-\frac{r^2h\Ddot{f}}{\xi^2}\right)\left(r\dot{f}+2\xi f^{1/2}\right)^2\right]_\Sigma \ .
\end{equation}

As a next step, replacing the second equation in ($\ref{28}$), ($\ref{31}$) and ($\ref{33}$) into ($\ref{35}$) provides 
\begin{equation}
 \label{37}    
    \frac{h(r_\Sigma)-r_\Sigma h'(r_\Sigma)-1}{r^2_\Sigma h(r_\Sigma)}+\frac{1}{4\xi^2}\frac{\dot{f}^2}{f}-\frac{\Ddot{f}}{\xi^2}=\frac{1}{2\xi}\frac{h'(r_\Sigma)}{h(r_\Sigma)}\frac{\dot{f}}{f^{1/2}} \ .
\end{equation}

Considering the static case for this expression, we reproduce the same result as Hernández and Núñez \cite{HernandezA}, that is,
\begin{equation}
  \label{38}      
     h'(r_\Sigma)=-\frac{1-h(r_\Sigma)}{r_\Sigma} \ .
\end{equation}
So, replacing ($\ref{38}$) in ($\ref{37}$), we have
\begin{equation}
   \label{38.1}
    \Ddot{f}-\frac{1}{4}\frac{\dot{f}^2}{f}-\xi\left(\frac{1-h(r_\Sigma)}{2r_\Sigma h(r_\Sigma)}\right)\frac{\dot{f}}{f^{1/2}}=0 \ ,
\end{equation}
which can be integrated to gives
\begin{equation}
   \label{41}
    \dot{f}=2a(f^{1/2}-f^{1/4}) \ ,
\end{equation}
where 
\begin{equation}
 \label{40} 
   a=-2\xi\left(\frac{1-h(r_\Sigma)}{r_\Sigma h(r_\Sigma)}\right) \ .
\end{equation}

Integrating the equation (\ref{41}), we find
\begin{equation}
  \label{42}
    t-t_0=\frac{f^{1/2}}{a}+\frac{2}{a}f^{1/4}+\frac{2}{a}\ln|f^{1/4}-1| \ ,
\end{equation}
with $t_0$ being an arbitrary constant.

Since $ f $ has values between $1$ and $0$, the previous equation can be rewritten as
\begin{equation}
  \label{43}
    t=\frac{f^{1/2}}{a}+\frac{2}{a}f^{1/4}+\frac{2}{a}\ln{(1-f^{1/4})} \ ,
\end{equation}
where we adopted $t_0=1$ without loss of generality.

So, when $-\infty < t\leq 1 $, it follows that $0<1-f^{1/4}\leq 1\ \Rightarrow \ 0 < f \leq 1 \ $, where $f\to 1$ when $t\to -\infty$.

Then, we get the field equations (\ref{30})-(\ref{33}) as a function of $f$ as follows:
\begin{equation}
  \label{45}    
    8\pi q= bh^{1/2}h'\left(\frac{f^{1/2}-f^{1/4}}{f^2}\right) \ ,
\end{equation}
\begin{equation}
 \label{46}    
    8\pi\rho= \frac{1-h-rh'}{fr^2}+3b^2h\left(\frac{f^{1/2}-f^{1/4}}{f}\right)^2 \ ,
\end{equation}
\begin{eqnarray}
 \label{47}    
    8\pi(P_r-\zeta\Theta)&=& \frac{h-rh'-1}{fr^2}+b^2h\left(\frac{f^{1/2}-f^{1/4}}{f}\right)^2\nonumber \\
    &-&b^2h\left(\frac{2f^{1/4}-1}{f^{3/4}}\right)\left(\frac{f^{1/2}-f^{1/4}}{f}\right) \ ,
\end{eqnarray}
\begin{eqnarray}
 \label{48}
    8\pi(P_\perp-\zeta\Theta)&=& \frac{h'^2-hh''}{2hf}+b^2h\left(\frac{f^{1/2}-f^{1/4}}{f}\right)^2\nonumber \\
    &-&b^2h\left(\frac{2f^{1/4}-1}{f^{3/4}}\right)\left(\frac{f^{1/2}-f^{1/4}}{f}\right) \ ,
\end{eqnarray}
where $b=a/\xi$.

Similarly, considering ($\ref{41}$), the expansion scalar in ($\ref{29}$) can be written as
\begin{equation}
 \label{49}    
    \Theta=3bh^{1/2}\left(\frac{f^{1/2}-f^{1/4}}{f}\right) \ .
\end{equation}

With the same change made in the equation above, the energy stored inside the hypersurface, expression ($\ref{34}$), takes the form
\begin{equation}
 \label{50}
    m=\frac{1}{2}\left[rf^{1/2}\left(1-h+\frac{(1-h)^2}{h}\frac{(f^{1/2}-f^{1/4})^2}{f}\right)\right]_\Sigma \ .
\end{equation}

Finally, considering ($\ref{38}$) and ($\ref{41}$), the equation ($\ref{36}$) for the luminosity is given as
\begin{eqnarray}
  \label{51}
    L_\infty&=&\frac{1}{2}\left\{\left[(1-h)^2\left(\frac{f^{1/2}-f^{1/4}}{f}\right)^2-(1-h)^2\left(\frac{2f^{1/4}-1}{f^{3/4}}\right)\left(\frac{f^{1/2}-f^{1/4}}{f}\right)\right]\right. \nonumber \\
    &\times& \left. \left[f^{1/2}+\frac{(1-h)}{h}(f^{1/2}-f^{1/4})\right]^2 \right\}_\Sigma \ .
\end{eqnarray}

\section{Solution for Anisotropic Fluid Static Spheres}

\ Hernández e Núñez \cite{HernandezA} investigated whether it is possible to obtain, at least in certain mass-radius ratio intervals of a spherically symmetric distribution of matter and static, physically reasonable anisotropic fluids satisfying an non-local equation of state. The choice of the metric, one  that coincides with (\ref{27}) for $\dot{f}=\Ddot{f}=0$, imposes the equation of state given by 

\begin{equation}
  \label{53} 
    P_r=\rho-\frac{2}{r^3}\int^r_0\Bar{r}^2\rho d\Bar{r}+\frac{C}{2\pi r^3} \ ,
\end{equation}
where $C$ is an arbitrary integration constant.
 
Supposing yet a density profile like the one originally proposed by Wyman \cite{Wyman}, as a special case of Tolman's solution VI \cite{Tolman}, given by
 
\begin{equation}
  \label{55} 
    \rho(r)=\frac{C}{8\pi} \frac{K(3+5Cr^2)}{(1+3Cr^2)^{5/3}}\ ,
\end{equation}
where $\rho$ is given in $s^{-2}$, C and K are constants to be determined, they found

\begin{equation}
   \label{60}
    h=1-\frac{2^{7/3}\gamma\delta^2}{(1+3\delta^2)^{2/3}} \ ,
\end{equation}
\begin{equation}
  \label{61}
    \rho=\frac{\gamma}{2^{2/3}\pi r_\Sigma^2}\frac{(5\delta^2+3)}{(3\delta^2+1)^{5/3}} \ ,
\end{equation}
\begin{equation}
 \label{62}
    P_r=\frac{\gamma}{2^{2/3}\pi r_\Sigma^2}\frac{(1-\delta^2)}{(3\delta^2+1)^{5/3}} \ ,
\end{equation}
\begin{equation}
  \label{63}
   P_\perp= \frac{\gamma}{2^{2/3}\pi r_\Sigma^2}\frac{2^{7/3}\gamma\delta^2(1+8\delta^2+3\delta^4)+(1+3\delta^2)^{2/3}(1-4\delta^2-\delta^4)}{(1+3\delta^2)^{8/3}[(1+3\delta^2)^{2/3}-2^{7/3}\gamma\delta^2]} \ ,
\end{equation}
where $\gamma=\frac{M_0}{r_\Sigma}$ is the mass-radius ratio and $\delta=\frac{r}{r_\Sigma}$ is defined in the range $0\leq \delta \leq 1 $, corresponding to $0\leq r\leq r_\Sigma$, respectively. Note that (\ref{60}) - (\ref{63}) already take into account the regularity and junction  with the Schwarzchild solution conditions

\ The result obtained for the tangential pressure differs from the one found in the article by Hernández and Núñez \cite{HernandezA}, due to a correction in the sign in the denominator in (\ref{63}). Consequently, we show below that there is a mass-radius ratio interval  for which  the spherically symmetrical and static distribution can be  constituted by a physically reasonable fluid in its entirety.

\begin{figure}[!ht]{}
  \label{Figura10}
  \subfloat[\label{4a}]{
    \includegraphics[width=0.47\hsize]
          {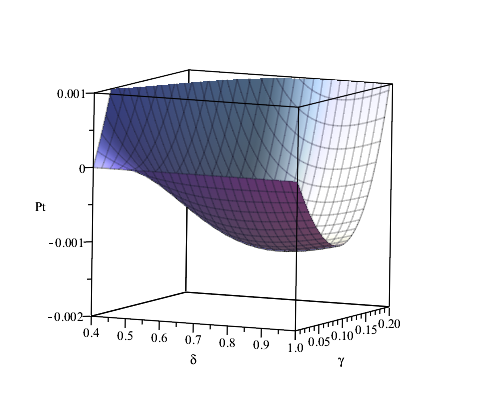}}
  \subfloat[ \label{4b}]{
    \includegraphics[width=0.38\hsize]
        {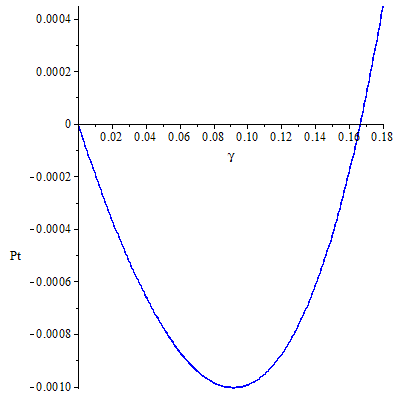}}\\
        
    \caption{In figure \ref{4a}, we have $P_t = P_\perp {r_\Sigma}^2$. While in figure \ref{4b} we have the limit for $\gamma$ imposed by the condition $P_\perp\geq 0$, considering $\delta = 1$.}    
\end{figure}

The figure \ref{4b} represents a cut in the figure \ref{4a} at $\delta=1$. We can verify that there is an interval of $\gamma$ corresponding to approximately $\gamma\leq0.170$, in which $P_\perp<0$. Negative pressures are not forbiden by general relativity, since the energy conditions are satisfied. In the next section we verify the limits imposed on $\gamma$ by the energy conditions.

\begin{figure}[H] 
\centering
\includegraphics[width=8.0cm]{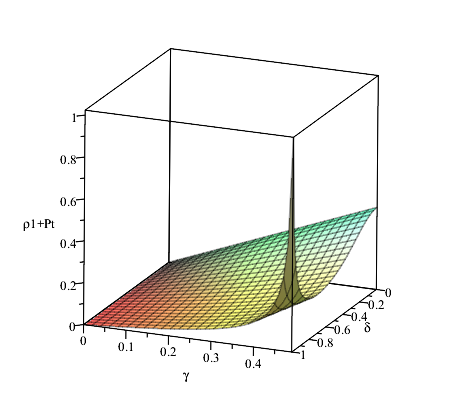}
\caption{In figure, we have $\rho1+P_t =(\rho+P_\perp)r_\Sigma^2$.}
\label{Figura5}
\end{figure}

\subsection{Energy conditions}

\ In general relativity theory, energy conditions are conditions that allow constraining energy-momentum tensors in order to select physically acceptable fluids. The main ones are divided into weak energy condition, dominant energy condition and strong energy condition. 

\ As we can see from the (\ref{61}) and (\ref{62}), the energy density and the radial pressure are always positive, assuring that the energy conditions $\rho\geq 0$ and $P_r\geq 0$ are satisfied. Furthermore, figure \ref{Figura5} shows that even with the small range of negative values for $P_\perp$ shown in figure \ref{4b}, also the energy condition $\rho+P_t\geq 0$ is also satisfied, all of them for $0\leq\delta\leq1$.

The energy condition $\rho-P_r\geq0$, through the equations ($\ref{61}$) and ($\ref{62}$), shows up also positive for every $\delta$ 
\begin{equation}
   \label{64}
    \rho-P_r=\frac{2^{1/3}\gamma}{\pi r_\Sigma^2(1+3\delta^2)^{2/3}} \geq 0\ ,
\end{equation}

\begin{figure}[!ht]{}
  
  \subfloat[\label{6a}]{
    \includegraphics[width=0.47\hsize]
     {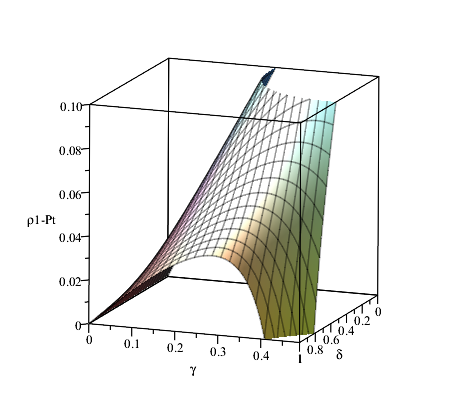}}
  \subfloat[ \label{6b}]{
    \includegraphics[width=0.37\hsize]
        {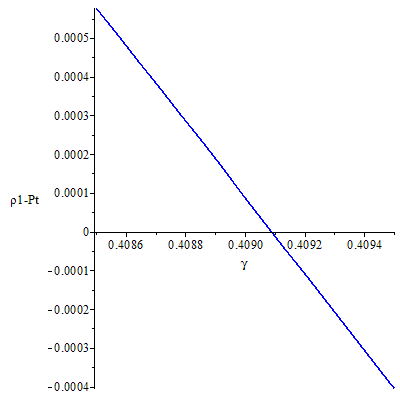}}\\
       
 \caption{ In figure \ref{6a} we have $\rho 1-P_t=(\rho -P_\perp)r_\Sigma ^2$. In the figure \ref{6b} we have the limit of $\gamma$ to satisfy the condition $\rho -P_\perp\geq 0$. }  
 \label{Figura6}
\end{figure}

\begin{figure}[H] 
\centering
\includegraphics[width=9.2cm]{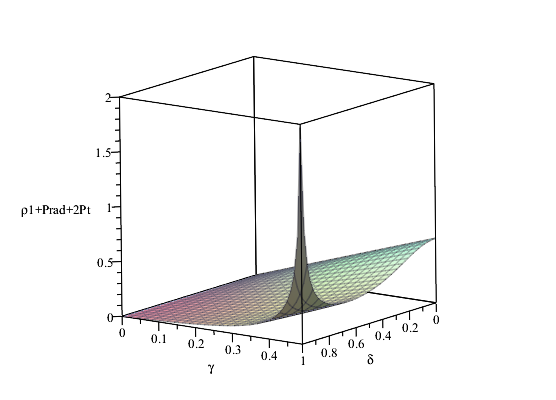}%
\caption{In figure, we have $\rho1+P_{\text{rad}}+2P_t=(\rho+P_r+2P_\perp)r_\Sigma^2$.}
\label{Figura7}
\end{figure}

The expressions for the energy conditions  $\rho-P_\perp\geq0$ and $\rho+P_r+2P_\perp\geq0$ are not obvious and their behavior can be easier studied graphically. According to the figure \ref{6a} we observe that there is  values for the mass-radius ratio in which the dominant energy condition $\rho-P_\perp\geq0$ is not satisfied. The figure \ref{6b} represents a section of the figure \ref{6a} in $\delta=1$ that shows this upper limit. So we should require $\gamma\leq 0.409$, approximately. Finally, the figure \ref{Figura7}, representing the strong condition $\rho+P_r+2P_\perp\geq0$, shows that it is always satisfied, indicating that our static model does not contain dark energy.
\ We therefore conclude that all acceptability conditions (regularity, hydrostatic equilibrium and energy conditions) are respected as long as the mass-radius ratio is in the range $0.170\leq\gamma\leq0.409$,
which in physical units is equivalent to $2.30\times 10^{26}$ Kg/m $\leq(\gamma)_{\text{fis}}\leq 5.51\times 10^{26}$ Kg/m, where $(\gamma)_{\text{fis}}=\frac{\gamma c^2}{G}$. These values contradict those found by Hèrnandez, Núñez \cite{HernandezA}, due to a correction in the sign of the tangential pressure, see the comment after the equation (\ref{63}). No restriction was found for the values of $\delta$. Thus, we have a spherically symmetrical distribution constituted by ordinary matter well behaved throughout its extension.

\section{Dynamic Evolution}

\ The previous session showed the initial static behavior of a spherically symmetric anisotropic fluid, where we determined the mass-radius ratio interval that satisfies all the acceptability conditions that make our model physically reasonable.

Now, let's follow the time evolution of this initial configuration through some important physical quantities  such as the total mass-energy, the luminosity seen by an observer at rest at infinity, the heat flux, expansion scalar, the effective surface temperature and the effective adiabatic index, during the collapse process.  
For this, considering the value for $\dot{f}$, equation ($\ref{41}$), we can obtain the instant of formation of the event horizon, that is,
\begin{equation}
  \label{66}
    f_H=16\gamma^4 \ .
\end{equation}

\ So, with $f_H$ and with the help of the equation ($\ref{40}$) we can rewrite the equation ($\ref{43}$) to obtain the instant of formation of the event horizon, $t_H$, which is
\begin{equation}
  \label{66.1}
    t_H=\frac{r_\Sigma}{\gamma}(2\gamma^2+2\gamma+\ln(1-2\gamma)) \ .
\end{equation}

\ We can rewrite the equations ($\ref{45}$), ($\ref{49}$), ($\ref{50}$) and ($\ref{51}$), considering (\ref{60}), respectively, as            
\begin{equation}
   \label{69}
    q=\frac{2^{4/3}\gamma^2\delta}{\pi r_\Sigma^2f^2}\frac{\left[1-\frac{2^{7/3}\gamma\delta^2}{(3\delta^2+1)^{2/3}}\right]^{1/2}(\delta^2+1)(f^{1/2}-f^{1/4})}{(2\gamma-1)(3\delta^2+1)^{5/3}} \ ,
\end{equation}
\begin{equation}
   \label{70}
    \Theta=\frac{6\gamma}{r_\Sigma f}\left[1-\frac{2^{7/3}\gamma\delta^2}{(3\delta^2+1)^{2/3}}\right]^{1/2}\frac{(f^{1/2}-f^{1/4})}{(1-2\gamma)} \ .
\end{equation}
\begin{equation}
 \label{67}    
    m=\gamma r_\Sigma f^{1/2}\left[1+\frac{2\gamma}{1-2\gamma}\frac{(f^{1/2}-f^{1/4})^2}{f}\right] \ ,
\end{equation}
\begin{eqnarray}
 \label{68}    
    L_\infty&=&2\gamma^2(f^{-1/4}-1)\left[1+\frac{2\gamma}{1-2\gamma}(1-f^{-1/4})\right]^2 \ .
\end{eqnarray}

\ The following results were obtained by substituting the equation (\ref{66.1}) in (\ref{67}), (\ref{69}) and (\ref{70}) in order to examine their behavior at the moment of the formation of the event horizon, through graphical analysis. This is possible because, for values of $f$$>$$f_H$, that is, for values of $f$ from the beginning of the collapse until moments before the formation of the event horizon, these functions are well behaved. For the equation (\ref{68}) this substitution will not add information, since $L_\infty(f=f_H)=0$, that is, in the formation of the horizon there is no detection of luminosity by the observer at infinity, as expected. As a result, our equations are
\begin{equation}
   \label{71}
    m_H=2r_\Sigma\gamma^2 \ ,
\end{equation}
\begin{equation}
 \label{72}
    q_H=\frac{2^{1/3}\delta(\delta^2+1)}{2^6\pi r_\Sigma^2\gamma^5}\frac{\left[1-\frac{2^{7/3}\gamma\delta^2}{(3\delta^2+1)^{2/3}}\right]^{1/2}}{(3\delta^2+1)^{5/3}} \ ,
\end{equation}
\begin{equation}
 \label{73}
    \Theta_H=-\frac{3}{4r_\Sigma \gamma^2}\left[1-\frac{2^{7/3}\gamma\delta^2}{(3\delta^2+1)^{2/3}}\right]^{1/2} \ .
\end{equation}

\ Making an analogy of our matter distribution with a star, we can consider the blackbody approximation, satisfying the Stefan-Boltzmann law, to obtain the analogue of its surface temperature. This law states that the total energy per unit time per unit area, $I$ (intensity), emitted by a blackbody in thermal equilibrium is proportional to the fourth power of its surface temperature, that is,
\begin{equation}
\label{74}
    I=\sigma T^4 \ , 
\end{equation}
where $T$ is the absolute temperature and $\sigma=\frac{1}{4}\omega c=5.6704\times10^{-8}\frac{W}{m^2 K^4}$ is the constant of Stefan-Boltzmann, with $w=\frac{8\pi^5 k_B^4}{15c^3h^3}$, and $c$ is the speed of light, $h$ is Planck's constant and $k_B$ is Boltzmann's constant. Taking into account the spherical symmetry and considering $R$ the radius of the surface of the star, we can obtain the intensity radiated on its surface as a function of the luminosity as
\begin{equation}
  \label{75} 
    I=\frac{L_\infty}{4\pi R^2} \ .
\end{equation}

\ The effective surface temperature of the star $(T_{\text{eff}})_\Sigma$ as measured by an observer at rest at infinity is the temperature that a blackbody would have to have in order to radiate the same amount of energy per meter square than the star \cite{Bohm}, that is, it is obtained through the equivalence of the equations (\ref{74}) and (\ref{75}), which results in
\begin{equation}
 \label{76} 
    \sigma (T_{\text{eff}}^4)_\Sigma=\frac{L_\infty}{4\pi R^2} \ \Rightarrow \ (T_{\text{eff}}^4)_\Sigma= \frac{L_\infty}{wc\pi R^2} \ .
\end{equation}

\ Considering the equality (\ref{19}), we can rewrite (\ref{76}) in geometric units ($G=c=1$) as,
\begin{equation}
\label{77}
  (T_{\text{eff}}^4)_\Sigma=\left(\frac{1}{w\pi C^2}\right)_\Sigma L_\infty \ .
\end{equation}

Thus, substuting (\ref{28}) and (\ref{68}) in the above equation, we obtain
\begin{equation}
\label{78}
   (T^4_{\text{eff}})= \frac{2\gamma^2}{\pi w r^2_\Sigma f}(f^{-1/4}-1)\left[1-\frac{2\gamma}{1-2\gamma}(f^{-1/4}-1)\right]^2 \ .
\end{equation}

Retrieving the physica
l units of the expression above, we find
\begin{equation}
\label{79}
   (T_{\text{eff}})= \alpha\gamma^{1/2}\left(\frac{f^{-1/4}-1}{f}\right)^{1/4}\left(1-\frac{2\gamma}{1-2\gamma}(f^{-1/4}-1)\right)^{1/2} \ ,
\end{equation}
where $\alpha=\left(\frac{15c^7h^3}{4\pi^6k_B^4G r_\Sigma^2}\right)^{1/4}$ has temperature dimension. Here, $G$ is gravitational constant. Considering that our matter distribution presents typical radii of a neutron star, that is, taking into account its minimum and maximum radii, 10 and 15 kilometers respectively, we can obtain the values of $\alpha$ in physical units. So $\alpha=5.2227\times10^{12}$K, for $r_\Sigma=10$Km and $\alpha=2.3212\times10^{12}$K, for $r_\Sigma=15$Km.

\ Besides, we can also obtain the effective adiabatic index, with the help of (\ref{46}) and (\ref{47}), that is,
\begin{equation}
\label{80}
    \Gamma_{\text{eff}}=\left[\frac{\partial (\ln P_r)}{\partial (\ln \rho)}\right]_{r=\text{cte}}=\left(\frac{\dot{P}_r}{P_r}\right)\left(\frac{\rho}{\dot{\rho}}\right) \ .
\end{equation}

\begin{figure}[!ht]{}
  \subfloat[$\iota\times\gamma$\label{10a}]{
    \includegraphics[width=0.4\hsize]
            {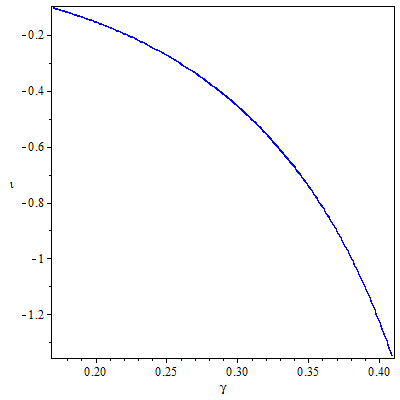}}\hfill
  \subfloat[$\gamma$=0.170 \label{10b}]{
    \includegraphics[width=0.4\hsize]
            {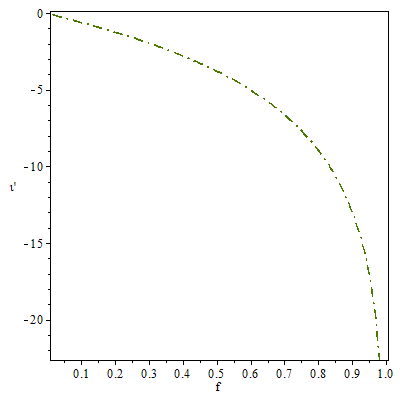}}\\
  \subfloat[$\gamma$=0.409 \label{10c}]{
    \includegraphics[width=0.4\hsize]
            {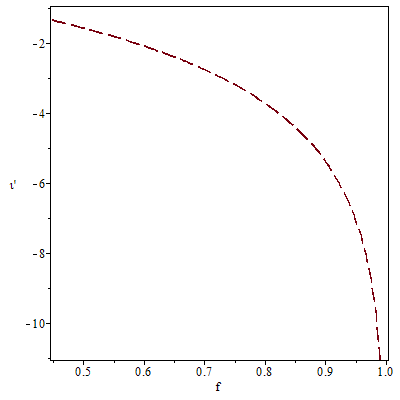}}\hfill
  
   \caption{ In figures, $\iota=\frac{t_H}{r_\Sigma}$ and $\iota'  =\frac{t}{r_\Sigma}$ . The function $f$ and $\gamma$ are dimensionless and the times $t_H$ and $t$ are in $s$.}
   \label{Figura10}
\end{figure}

\begin{figure}[!ht]{}
  \subfloat[$\mu \times \gamma$ \label{11a}]{
    \includegraphics[width=0.4\hsize]
            {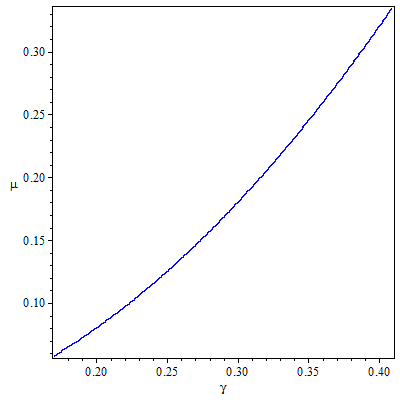}}\hfill
  \subfloat[$\gamma$=0.170 \label{11b}]{
    \includegraphics[width=0.4\hsize]
            {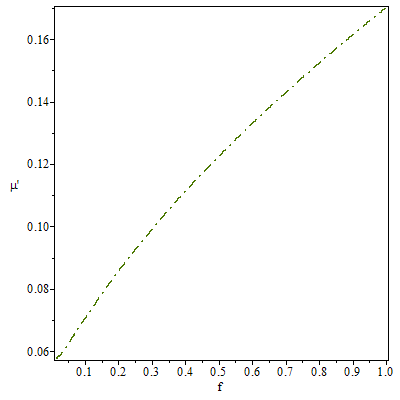}}\\
  \subfloat[$\gamma$=0.409 \label{11c}]{
    \includegraphics[width=0.4\hsize]
            {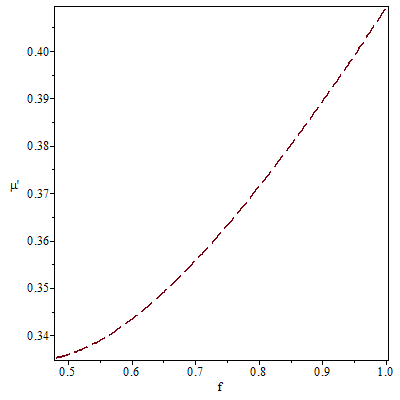}}\hfill
  
  \caption{In figures, $\mu=\frac{m_H}{r_\Sigma}$ and $\mu'=\frac{m}{r_\Sigma}$. The masses $m_H$ and $m$ are in $s$, the function $f$ and $\gamma$ are dimensionless. The mass loss in 11b is 66\%, whereas in 11c it was a loss of 18.2\%.}
  \label{Figura11}
\end{figure}

\begin{figure}[!ht]{}
  \subfloat[$\gamma$=0.170 \label{12a}]{ 
    \includegraphics[width=0.4\hsize]
            {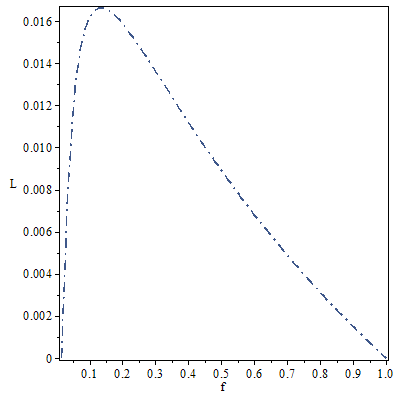}}\hfill
  \subfloat[$\gamma$=0.3\label{12b}]{
    \includegraphics[width=0.4\hsize]
            {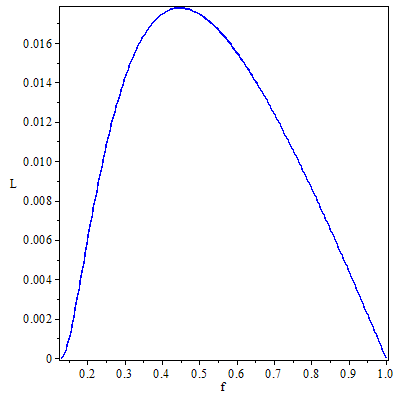}}\\
  \subfloat[$\gamma$=0.409 \label{12c}]{
    \includegraphics[width=0.4\hsize]
            {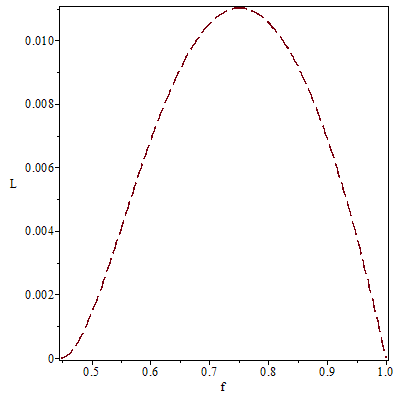}}\hfill
 
 \caption{The luminosity L and the function $f$ are dimensionless.}
 \label{Figura12}
\end{figure}

\begin{figure}[!ht]{}
  \subfloat[Q$\times\gamma\times\delta$ \label{13a}]{ 
    \includegraphics[width=0.45\hsize]
            {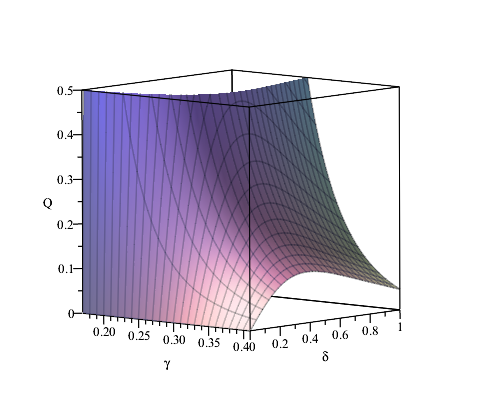}}\hfill
  \subfloat[$\gamma=0.170$ \label{13b}]{
    \includegraphics[width=0.45\hsize]
            {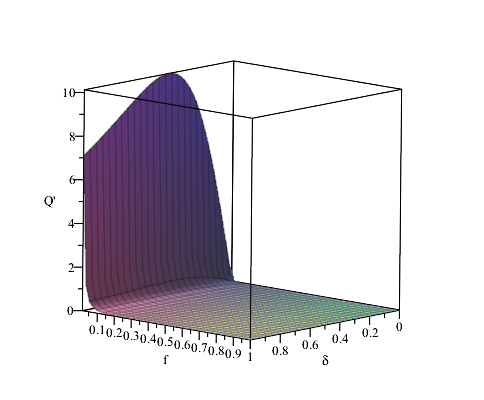}}\\
  \subfloat[$\gamma=0.409$ \label{13c}]{
    \includegraphics[width=0.45\hsize]
            {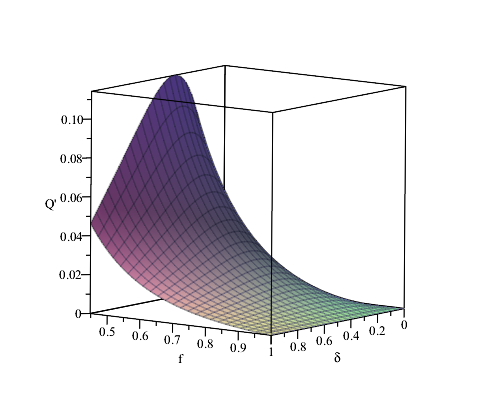}}\hfill
 
  \caption{In figures, $Q=(q_H) r_\Sigma^2$ and $Q'=qr_\Sigma^2$. The heat fluxes $q$ and $q_H$ are in $s^{-2}$, the function $f$ and $\gamma$ are dimensionless.}
  \label{Figura13}
\end{figure}

\begin{figure}[!ht]{}
  \subfloat[$\theta\times\gamma\times \delta$ \label{14a}]{ 
    \includegraphics[width=0.45\hsize]
            {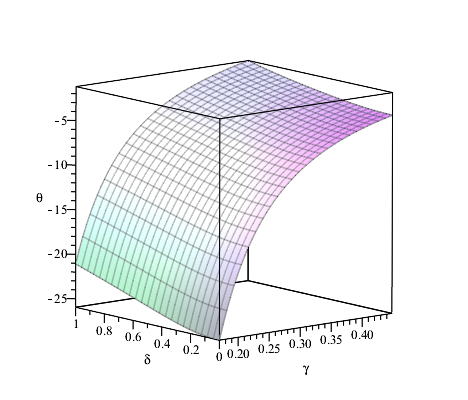}}\hfill
  \subfloat[$\gamma=0.170$ \label{14b}]{
    \includegraphics[width=0.45\hsize]
            {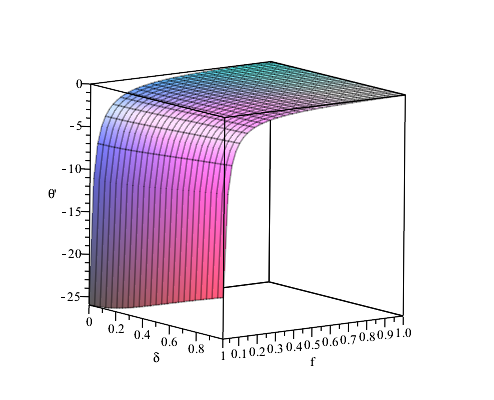}}\\
  \subfloat[$\gamma=0.409$ \label{14c}]{
    \includegraphics[width=0.45\hsize]
        {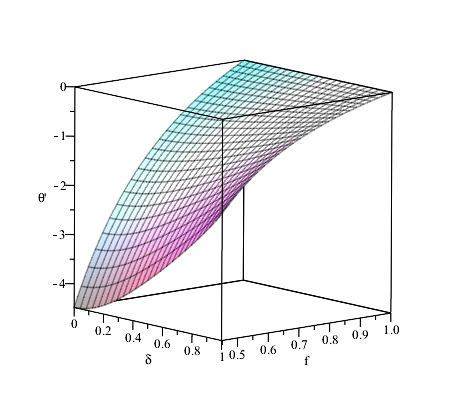}}\hfill
  
  \caption{In figures, $\theta=(\Theta_H) r_\Sigma$ and $\theta'=\Theta r_\Sigma$. The scalars $\Theta_H$ and $\Theta$ are in $s^{-1}$, the function $f$ and $\gamma$ are dimensionless.}
  \label{Figura14}
\end{figure}

\begin{figure}[!ht]{}
  \subfloat[$\gamma$=0.170 \label{15a}]{ 
    \includegraphics[width=0.4\hsize]
            {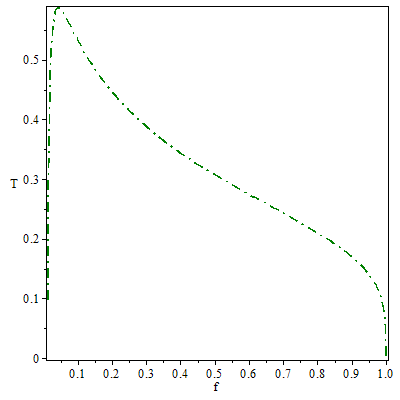}}\hfill
  \subfloat[$\gamma$=0.3 \label{15b}]{
    \includegraphics[width=0.4\hsize]
            {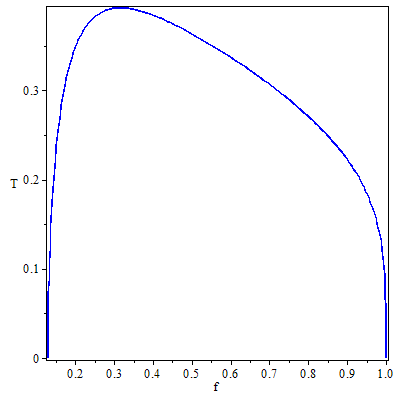}}\\
  \subfloat[$\gamma$=0.409 \label{15c}]{
    \includegraphics[width=0.4\hsize]           {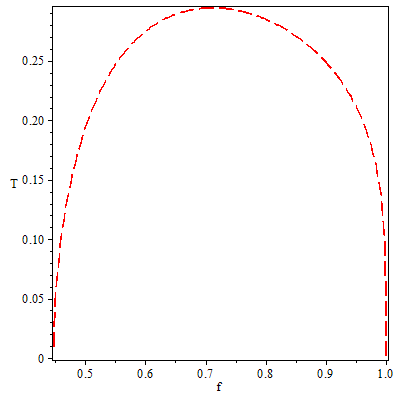}}

 \caption{In figures, $T=\frac{T_{\text{eff}}}{\alpha}$. The effective surface temperature is given in K and the function $f$ is dimensionless.}
 \label{Figura15}
\end{figure}

 This coefficient is obtained when we consider that the fluid that composes the star behaves (or has similar characteristics) as an ideal fluid, whose adiabatic index is given by the ratio between the specific heats \cite{Rezola}, being a particular case where the total  pressure inside the star is made up of just the pressure of the fluid \cite{ChandrasekharC}. In the study of stellar stability, the adiabatic index translates the rigidity of the equation of state, for a given energy density and is defined for adiabatic processes. It is often used to evaluate the stability of the star, where the value 4/3 is considered its lower limit, in order to minimally ensure its stability \cite{Casali, Moustakidis, Esculpi}. Here, we consider an effective adiabatic index, since we are not in the adiabatic regime. Although this measurement is not a sufficient condition to guarantee the stability of the \cite{PretelB} star, it is usually used as a first approach, and we will not do otherwise here.
\ With these results, we can perform a graphical analysis of these quantities during gravitational collapse, with special attention to the moment of formation of the event horizon. The figure \ref{Figura10} shows the relationship between $t_H$ and $\gamma$ and also between $t$ and $f$, based on the equations (\ref{66.1}) and (\ref{43}) . We can observe in the figure \ref{10a} that, as $\gamma$ increases, we consequently have a decrease in the time of formation of the event horizon. This means that stars with increasing mass-to-radius ratios collapse faster. This makes sense, because if we fix the radius of the star, we would expect the gravitational force to be greater on increasingly massive objects, so that they collapse{Rezzolla} more quickly. The figures \ref{10b} and \ref{10c}, represent the time $t$ of the lowest and highest value for $\gamma$, respectively, from the beginning of the collapse, $f=1$, to the formation of the black hole, coaxing us to look at these graphs from right to left. In the following graphs we will also see the behavior of the physical quantities for both the lower and upper limits of $\gamma$.

\ In the figure \ref{Figura11} we observe the behavior of the total mass-energy of the star in relation to $\gamma$ and $f$, given by the expressions (\ref{71}) and (\ref{69}). The figure \ref{11a} shows all the values of $m_H$, equation (\ref{71}), at the moment of formation of the event horizon. In it we notice that, at the end of the collapse, the total mass-energy is larger for increasing values of $\gamma$. Figures \ref{11b} and \ref{11c} show the behavior of mass loss $m$, equation (\ref{69}), in the form of radiation, for both limits of $\gamma$ already established above.

\ Then, the images of the figure \ref{Figura12} reveal the luminosity detected by an observer at infinity with the help of the equation ($\ref{68}$). The observer would see a sudden increase in the star's brightness and then an abrupt decrease in light until the moment of the formation of the event horizon. In a contradictory way, the luminosity peak is higher for intermediate values of $\gamma$. We can also see that this sudden decrease in luminosity is more pronounced for stars with a lower mass-to-radius ratio.

\ In the images of the figure \ref{Figura13} we have the behavior of the heat flow considering the equations (\ref{72}) and (\ref{69}). With the figure \ref{13a} we notice that, at the moment of formation of the event horizon, $q_H$ presents smaller and smaller peaks as the mass-radius ratio increases. The graphs of figures \ref{13b} and \ref{13c} are inserted to emphasize this result and show the development of the respective limit values of $\gamma$ along the collapse considering the expression (\ref{69}) .

\  Meanwhile, the figure \ref{Figura14} shows the graphs of the expansion scalar obtained through (\ref{73}) and (\ref{70}). It can be seen that, in the figure \ref{14a}, $\Theta_H$ decreases more and more the smaller the value of $\gamma$. Similarly to the heat flux, the figures \ref{14b} and \ref{14c} were inserted to highlight this result and show its development along the collapse considering the equation (\ref{70}).

\ In the figure \ref{Figura15} we find a behavior similar to luminosity, that is, the effective temperature on the surface of the star measured by an observer at infinity suddenly increases at a given moment to a maximum value and then decreases rapidly until it becomes a black hole. Furthermore, the peak for $T_{\text{eff}}$ assumes a larger value for the smaller $\gamma$. 
 
\begin{figure}[!ht]{}
  \subfloat[$\delta$=1 \label{18.1a}]{ 
    \includegraphics[width=0.45\hsize]
            {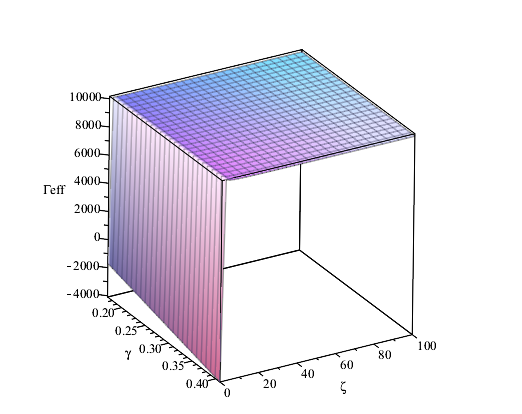}}\hfill
  \subfloat[$\delta$=1 \label{18.1b}]{
    \includegraphics[width=0.45\hsize]
            {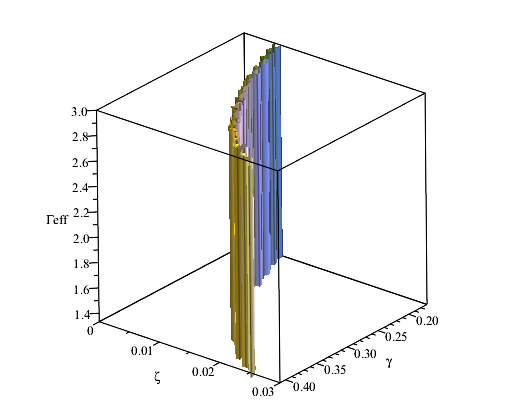}}\\
 
 \caption{$\Gamma_{\text{eff}}$ becomes unstable for small values of $\zeta$ at the edge of the distribution. }
 \label{Figura18.1}
\end{figure}

\begin{figure}[!ht]{}
  \subfloat[$\delta$=0 \label{18.2a}]{ 
    \includegraphics[width=0.45\hsize]
            {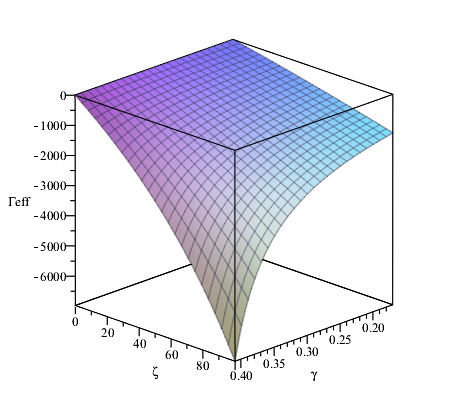}}\hfill
  \subfloat[$\delta$=0 \label{18.2b}]{
    \includegraphics[width=0.45\hsize]
            {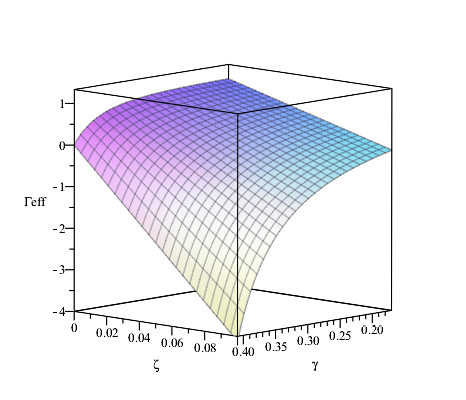}}\\

\caption{$\Gamma_{\text{eff}}$ is always unstable for any value of $\zeta$ in the center of the distribution. }
\label{Figura18.2}
\end{figure}

 In the figures \ref{Figura18.1}, and \ref{Figura18.2} we show how the effective adiabatic index can be modified by the volume viscosity and how it relates to the different models as a function of the mass-radius ratio. The figures \ref{Figura18.1} and \ref{Figura18.2} show the evolution of the effective adiabatic index as a function of the mass-radius ratio and the volume viscosity coefficient, both on the surface, $\delta=1$ , and in the center, $\delta=0$, for $f=0.9999$ (very close to the initial static case). Although in this work $\zeta$ is being considered a constant, strictly it does not depend only on the fluid that composes the star, but also on quantities such as pressure and temperature, in order to depend on the process. Therefore, we are analyzing the behavior of $\Gamma_{\text{eff}}$ as a function of $\zeta$ and $\gamma$, at an instant very close to the initial instant, in two extreme regions of the star. The graph of figure \ref{18.1a} reveals the behavior of the effective adiabatic index for an interval $0\leq\zeta\leq 100$ on the surface, and we notice a sudden decline in the effective adiabatic index as $\zeta$ approaches zero. This boundary between stability and instability is best seen in the graph of figure \ref{18.1b}, which shows a cut at $4/3\leq\Gamma_{\text{eff}}\leq 3$ and $0\leq\zeta\leq 0.03$, from the previous figure. In it, we can see that on the right side of the surface, we find the points where the star is stable, while on the left side of the surface we have the points where the star is unstable. It indicates how stability depends on combinations between $\gamma$ and $\zeta$. The figure \ref{Figura18.2} expresses the same general behavior for $\Gamma_{\text{eff}}$, however, calculated at the center of the star. The figure \ref{18.2a} provides the behavior of the effective adiabatic index again in the range $0\leq\zeta\leq 100$, which apparently indicates instability for any value of the volume viscosity coefficient. As the range for $\zeta$ is very large, we cannot carefully observe whether there is stability at small values. So, the figure \ref{18.2b}, which represents a cut of the previous figure in $0\leq\zeta\leq 0.1$, points out that the star is always unstable for small values of $\zeta$ and for any $\gamma$ of our range. This result is interesting, as it means that for any given volume of viscosity, the star's collapse is its natural fate. Therefore, in this case, the star does not collapse only because of the chosen metric, but also because it represents an initial static but unstable configuration.




\section{Energy conditions for the dynamic case}

\ By examining the acceptability conditions of the model for the static case we found an interval for the mass-radius ratio where all energy conditions were satisfied. Now, we will examine the dynamic case with the intention of verifying if the model will consist of a physically reasonable fluid under the same ranges of $\gamma$, or if these will be modified during the collapse process. For this, we will base ourselves on the articles by Kolassis et al \cite{Kolassis} and Veneroni e da Silva \cite{Veneroni}, generalizing the energy conditions for an anisotropic fluid with viscosity and heat flux.

\ As the energy-momentum tensor given in ($\ref{5}$) is not diagonal, we need to perform its diagonalization to obtain its eigenvalues $\lambda$ which are the roots of the equation
\begin{equation}
 \label{81}    
    |T^-_{\alpha\beta}-\lambda g_{\alpha\beta}|=0 \ .
\end{equation}
In this case, it is with these eigenvalues that we will obtain the energy conditions.

\begin{equation}
 \label{82}
    P_1=P_r-\zeta\Theta=\frac{1}{8\pi}\left(\frac{h-rh'-1}{r^2f}+\frac{h}{4\xi^2}\frac{\dot{f}^2}{f^2}-\frac{h}{\xi^2}\frac{\Ddot{f}}{f}\right) \ ,
\end{equation}
\begin{equation}
 \label{83}
    P_2=P_\perp-\zeta\Theta=\frac{1}{8\pi}\left(\frac{h'^2-hh''}{2hf}+\frac{h}{4\xi^2}\frac{\dot{f}^2}{f^2}-\frac{h}{\xi^2}\frac{\Ddot{f}}{f}\right) \ ,
\end{equation}
\begin{equation}
 \label{84}    
    \overline{q}=qB=\sqrt{\frac{f}{h}}q \ .
\end{equation}

\ In this way, the determinant of (\ref{81}) will be written in terms of these quantities, that is,
\begin{equation}
  \label{84}
    (P_2-\lambda)(P_2-\lambda)[(\rho+\lambda)(P_1-\lambda)-\overline{q}^2]=0 \ .
\end{equation}
This results in the following roots:
\begin{eqnarray}
 \label{85}    \lambda_0=-\frac{1}{2}(\rho-P_1+\Delta) \ ,
\end{eqnarray}
\begin{eqnarray}
  \label{86}
    \lambda_1 = -\frac{1}{2}(\rho-P_1-\Delta) \ ,
\end{eqnarray}
\begin{eqnarray}
  \label{87}
    \lambda_2=\lambda_3
    = P_2 \ ,
\end{eqnarray}
where $\Delta^2$ is the discriminant of the equation, which is given by
\begin{equation}
  \label{88}
    \Delta^2=(\rho+P_1)^2-4\overline{q}^2 \ ,
\end{equation}
where $\Delta^2\geq0$ gives real solutions and $\Delta^2<0$ gives imaginary solutions.

\ This process has given us the desired roots which will then provide the proper energy conditions.

\subsection{Weak energy conditions}

\ Weak energy conditions are satisfied for the following relationships between the eigenvalues:
\begin{equation}
  \label{89}
    -\lambda_0\geq0 \ , 
\end{equation}
\begin{equation}
   \label{90}
    -\lambda_0+\lambda_i\geq0\ .
\end{equation}
The first weak energy condition is obtained from the expressions ($\ref{85}$) and ($\ref{89}$), which leads us to the result
\begin{equation}
 \label{91}    
    \rho -P_1+\Delta\geq0 \ .
\end{equation}

\ Analogously, considering ($\ref{89}$), ($\ref{85}$) and ($\ref{86}$), for $i=1$, we obtain
\begin{equation}
 \label{92}    
    \Delta\geq0 \ .
\end{equation}

\ Finally, through ($\ref{89}$), ($\ref{85}$) and ($\ref{87}$), for $i=2.3$,
 we found
 \begin{equation}
 \label{93}    
     \rho-P_1+2P_2+\Delta\geq0 \ .
 \end{equation}

\subsection{Dominant energy conditioshows the relationship betweenns}

\ The prevailing power conditions are equivalent to
\begin{equation}
  \label{94}
    -\lambda_0\geq0 \ ,
\end{equation}
\begin{equation}
 \label{95}    
    \lambda_0\leq\lambda_i\leq-\lambda_0 \ .
\end{equation}

\ Following the same steps described above, through ($\ref{94}$), we conclude that the first dominant energy condition is the same as the one obtained ($\ref{91}$), that is,
\begin{equation}
  \label{96} 
    \rho -P_1+\Delta\geq0 \ .
\end{equation}

\ Now, considering the equations ($\ref{95}$), ($\ref{85}$) and ($\ref{86}$), for $i=1$, we obtain the following inequality:
\begin{eqnarray}
 \label{97}
   0\leq\Delta)\leq(\rho -P_1+\Delta) \ .
\end{eqnarray}

\ Through this we can remove other inequalities. Are they:
\begin{equation}
 \label{98}
    \Delta\geq0 \ ,
\end{equation}
\begin{equation}
 \label{99}
    \rho -P_1\geq0 \ .
\end{equation}

\ Likewise, considering ($\ref{95}$), ($\ref{85}$) and ($\ref{87}$), for $i=2.3$, we have
\begin{equation}
 \label{100}    
    \rho-P_1+2P_2+\Delta\geq0 \ ,
\end{equation}
\begin{equation}
\label{101}
    \rho-P_1-2P_2+\Delta\geq0 \ .
\end{equation}

\subsection{Strong energy conditions}

\ Finally, we have the strong energy conditions, where
\begin{equation}
  \label{102}
    -\lambda_0+\sum_i \lambda_i\geq0 \ ,
\end{equation}
\begin{equation}
  \label{103} 
    -\lambda_0+\lambda_i\geq0 ,
\end{equation}
where $i=1,2,3$.

 \ First, let's check ($\ref{102}$) with the substituti
 \begin{equation}
   \label{104}
     2P_2+\Delta\geq0 \ .
 \end{equation}
 
 \ Next, we can see that the expressions ($\ref{90}$) and ($\ref{103}$) are the same and we obtain the same results as in the case of weak energy, that is,
 \begin{equation}
   \label{105}
     \Delta\geq0 \ , 
 \end{equation}
 \begin{equation}
    \label{106}
       \rho-P_1+2P_2+\Delta\geq0 \ .
 \end{equation}
 
 \ As a way to better visualize, the table \ref{tab:tabela2} summarizes all these energy conditions to which these inequalities belong.
 \begin{center}
\begin{table}
\caption{Energy conditions. 
\label{tab:tabela2}}%
\begin{tabular}{ |c|c|c| } 
 \hline
 $\Delta\geq0$ & WEC/ DEC/ SEC  \\ 
 $\rho-P_1+\Delta\geq0$ & WEC/ DEC  \\
 $\rho-P_1\geq0$ & DEC \\
 $\rho-P_1-2P_2+\Delta\geq0$ & DEC  \\
 $\rho-P_1+2P_2+\Delta\geq0$ & WEC/ DEC/ SEC  \\
 $2P_2+\Delta\geq0$ & SEC  \\
 \hline
\end{tabular}
\end{table}
\end{center}

\begin{figure}[H]
\centering
\includegraphics[width=7.0cm]{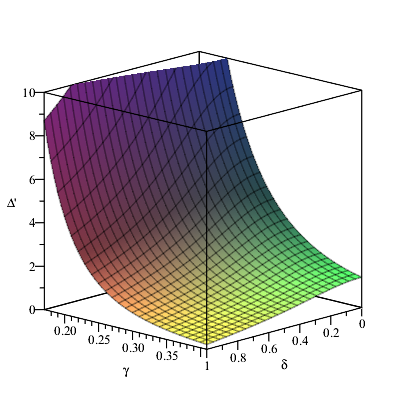}%

\caption{In figure, $\Delta'=\Delta r_\Sigma^2$. Function $f$ is dimensionless, and function $\Delta$ is in $s^{-2}$.}
\label{Figura18}
\end{figure}

\begin{figure}[H] 
\centering
\includegraphics[width=7.0cm]{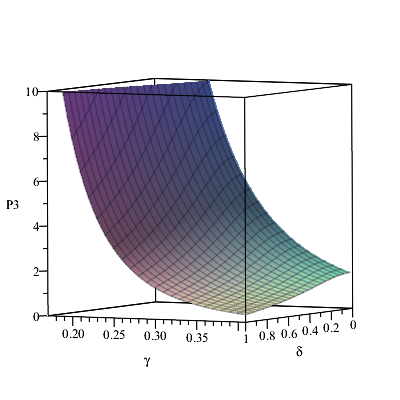}%

\caption{In figure, $P_3=(\rho-P_1+\Delta)r_\Sigma^2$. Function $f$ is dimensionless, and function $\rho-P_1+\Delta$ is in $s^{-2}$.}
\label{Figura19}
\end{figure}

\begin{figure}[H] 
\centering
\includegraphics[width=7.0cm]{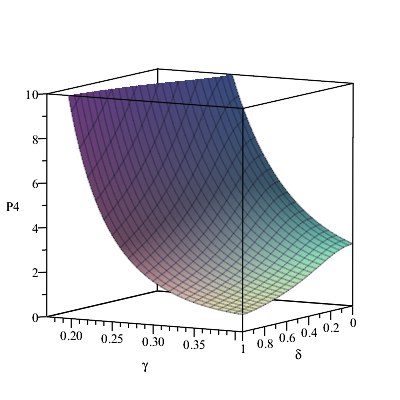}%

\caption{In figure, $P_4=(\rho-P_1+2P_2+\Delta)r_\Sigma^2$. Function $f$ is dimensionless, and function $\rho-P_1+2P_2+\Delta$ is in $s^{-2}$.}
\label{Figura20}
\end{figure}

\begin{figure}[H] 
\centering
\includegraphics[width=7.0cm]{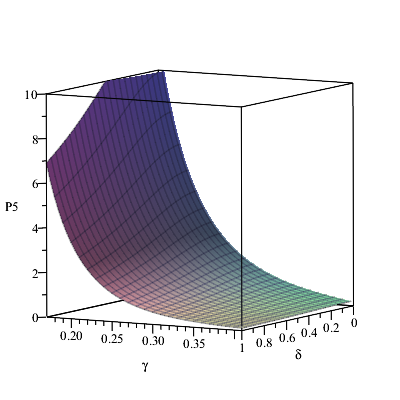}%

\caption{In figure, $P_5=(\rho-P_1)r_\Sigma^2$. Function $f$ is dimensionless, and function $\rho-P_1$ is in $s^{-2}$.}
\label{Figura21}
\end{figure}

\begin{figure}[H]
\centering
\includegraphics[width=7.0cm]{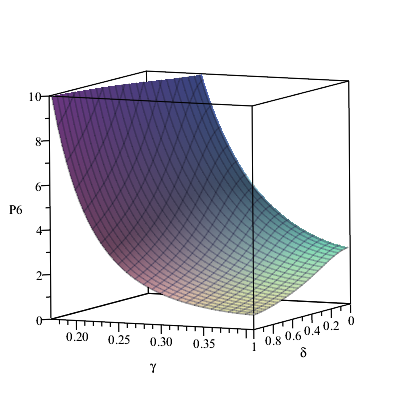}%

\caption{In figure, $P_6=(2P_2+\Delta)r_\Sigma^2$. Function $f$ is dimensionless, and function $2P_2+\Delta$ is in $s^{-2}$.}
\label{Figura22}
\end{figure}

\begin{figure}[H] 
\centering
\includegraphics[width=7.0cm]{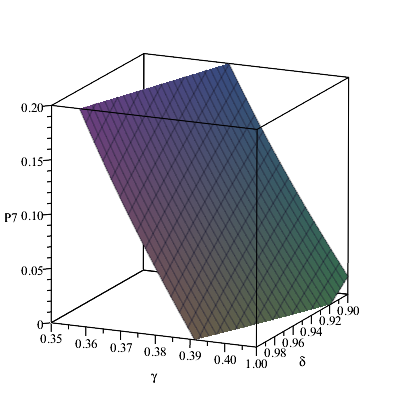}%

\caption{In figure, $P_7=(\rho-P_1-P_2+\Delta)r_\Sigma^2$. Function $f$ is dimensionless, and function $\rho-P_1-P_2+\Delta$ is in $s^{-2}$.}
\label{Figura23}
\end{figure}

\begin{figure}[H] 
\centering
\includegraphics[width=7.0cm]{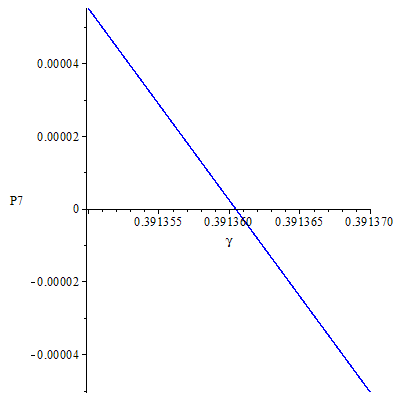}%

\caption{Limit of $\gamma$ to satisfy the condition $\rho -P_1-P_2+\Delta\geq 0$. Function $f$ is dimensionless, and function $P_7$ is in $s^{-2}$.}
\label{Figura24}
\end{figure}

 \subsection{Graphic representation of energy conditions}
 
 \ To perform the graphical analysis, we will adopt the same idea pointed out in Section IV, substituting the equation (\ref{66}) in the energy conditions to point out their behavior at the moment of formation of the event horizon.

\ Thus, with the help of the table \ref{tab:tabela2}, it is observed through the figures \ref{Figura18} and \ref{Figura21} that the first and third inequalities are satisfied and, consequently, the second constraint is also satisfied. is, as shown in the figure \ref{Figura19}. Analogously, the sixth condition is satisfied as seen in figure \ref{Figura22}. Therefore, the union of this with the third inequality leads to the conclusion that the fifth condition is fulfilled, as we can see in Figure \ref{Figura20}. It remains only to check the fourth constraint, represented in the figure \ref{Figura23}, where we can see that the upper bound for $\gamma$ is now smaller than the one obtained in the static case.   

\ Then, the figure \ref{Figura24} shows a clipping of the previous figure in $\delta$=1 revealing the value of the mass-radius ratio from which this dominant energy condition is violated. Therefore, for values of $\gamma\leq$ $0.3905$, approximately, this condition is satisfied. We therefore now have a new range that makes the dynamic model physically possible, where $0.170\leq\gamma\leq0.390$. In physical units this interval is equivalent to $2.30\times 10^{26}$ Kg/m $\leq\gamma\leq 5.26\times10^{26}$ Kg/m.
 
\section{Conclusion}

\ In this work, a study of the gravitational collapse was made for a spherically symmetric radiant distribution, which undergoes dissipation in the form of radial heat flux and emits zero radiation at the surface.

\ In order to study the behavior of physical quantities along the collapse, we introduce a time-dependent solution in the interior space-time metric, in order to follow the temporal behavior of these quantities from the initial static configuration to the formation of the event horizon, when the matter distribution becomes a black hole.

In this model, our analyzes were performed in terms of the $f(t)$ function, since it was not possible to explicitly obtain the temporal function present in the metric. However, as the function $f$ varies between 1 and 0 along the collapse, it was possible to examine all relevant physical quantities in terms of this function, where $f\to 1$ represents the initial static configuration.

\ For the static case, Einstein's equations for an anisotropic fluid at pressures naturally lead to a nonlocal equation of state, making it possible to obtain an exact analytical solution. This initial static distribution was taken from Tolman's VI solution, where the fulfillment of acceptability conditions restricted the values of the mass-radius ratio, generating an interval given by $2.30\times10^{26}Kg/m\leq (\gamma)_ {\text{fis}} \leq 5.51\times10^{26}Kg/m$. Remembering that this interval differs from the one obtained by Hernández \& Núñez, since we corrected the sign of the tangential pressure.

\ During the collapse process, when the nonlocal equation of state is no longer satisfied, we have seen that stars with higher mass-to-radius ratios collapse faster. Then, there will be a greater loss of mass for objects with smaller and smaller mass-radius ratios, until they become a black hole. This mass loss occurs in the form of radiation and, as we have observed, the peak in heat flux is greater for smaller and smaller values of the mass-radius ratio, indicating a greater mass loss in the form of radiation for these stars moments before become a black hole. We also notice that the luminosity starts from zero, grows rapidly until it reaches a peak and then suddenly decreases to the initial value. In this way, a sudden brightness will be detected by the observer, which will decrease abruptly just before the star becomes a black hole. A behavior similar to that of luminosity is obtained for the effective surface temperature.

\ On the other hand, the study carried out for the energy conditions for the dynamic case led to an even greater restriction for the upper limit of $\gamma$, if we want the star to satisfy all conditions during the entire collapse process, which is now in the interval $2.30\times 10^{26}Kg/m\leq(\gamma)_{\text{fis}}\leq 5.26\times10^{26}Kg/m$. So, now, our model leads to configurations that, in fact, can describe a spherically symmetrical distribution composed of a physically reasonable fluid in all its extension, from its initial static configuration to the formation of the event horizon. It is interesting to point out that such mass-radius ratio values differ from those obtained in other works. For example, in \cite{Veneroni} a very small range was found, $5.0625\times 10^{26}Kg/m\leq(\gamma)_{\text{fis}}\leq 5.906525\times 10^{26}Kg/m$, which is very close to the upper limit obtained here. In \cite{Pretel} an even larger range was reached, $0<(\gamma)_{\text{fis}}\leq5.2866\times10^{26}Kg/m$. Such ranges differ significantly due to the density profile used by each. In the first, the profile of Gokhroo \& Mehra was adopted and, in the second, the profile of Tolman's IV solution was used. Therefore, the density profile used has a great influence on the results obtained, taking into account that the same temporal dependence on the metric was used in all the works mentioned above.
We can also present these limits in terms of the number of solar masses, to facilitate comparison with some observations. If we consider again that our matter distribution represents a neutron star, with its typical values for the minimum and maximum radius, that is, $r_\Sigma=10Km$ and $r_\Sigma=15Km$, respectively, we get $1.15M_\odot\leq M_0\leq 2.75M_\odot$, for $r_\Sigma=10Km$, and $1.72M_\odot\leq M_0\leq 4.13M_\odot$, for $r_\Sigma=15Km$, for the initial compact object with $0.170\leq\gamma\leq0.409$, while for the black hole formed in the collapse we would have $0.39M_\odot\leq M_{\text{BH}}\leq2.05M_\odot$ (where $M_{\text{BH}}$ is the mass of the black hole formed), for $r_\Sigma=10Km$, and $0.58M_\odot\leq M_{\text{BH}}\leq 3.07M_ \odot$, for $r_\Sigma=15Km$ considering the range $0.170\leq\gamma\leq 0.390$.  Although modern estimates, using both binary neutron star and neutron star–black hole binaries, obtainned a broad neutron star mass distribution extending from $1.2{ M }_ {\odot }$ to $2.0{ M }_ {\odot }$ \cite{Abbott2022}, recent observations \cite{AbbottC} suggest the existence of a compact object of about $2.50-2.67M_\odot$, which could be identified with either a massive neutron star or a low-mass black hole. More recently, and even from the detection of gravitational waves from compact objects, evidence has been found of binary neutron star-black hole systems with masses of their components given by $8.{9}{ M }_ {\odot }$ - $1.{ 9}{ M }_ {\odot }$ and $5.{7}{ M }_ {\odot }$ - $1.{5}{M}_{\odot }$, for named events GW200105 and GW200115, respectively \cite{Abbott2021}. Therefore, the neutron mass range obtained here is reasonably in agreement with the observations.

\section*{Acknowledgments}

The financial assistance from Conselho Nacional de Desenvolvimento Científico (CNPq), Fundação Carlos Chagas Filho de Amparo à Pesquisa do Estado do Rio de Janeiro (FAPERJ), and Coordenação de Aperfeiçoamento de Pessoal de Nível Superior (CAPES) are gratefully acknowledged.

\section{Appendix: Junction condition}

The requirement of the matching between the inner (matter distribution) and outer (radiation zone around the distribution) solution across the $\Sigma$ hipersurface is called as the junction condition problem. This condition imposes continuity of the first and second fundamental forms, the first being the metric and the second the extrinsic curvature, such as the conditions established by Israel \cite{Israel}. So, when we approach $\Sigma$ through interior or exterior spacetime we must demand that
\begin{equation}
  \label{14}
    (ds^2_-)_\Sigma=(ds^2_+)_\Sigma=ds^2_\Sigma
\end{equation}
where $()_\Sigma$ means the value of $()$ over $\Sigma$. Thus, the equations when taken on the hypersurface $\Sigma$ obtained from the inner and outer spacetimes are $\chi_\pm^\alpha=\chi_\pm^\alpha(\xi^i)$. The extrinsic curvature of $\Sigma$ is given by \cite{Eisenhart} 
\begin{equation}
  \label{15} 
    K_{ij}^\pm\equiv -\eta^\pm_\alpha \frac{\partial^2\chi_\pm^\alpha}{\partial\xi^i\partial\xi^j}-\eta_\alpha^\pm\Gamma_{\beta\gamma}^\alpha\frac{\partial \chi_\pm^\beta}{\partial\xi^i}\frac{\partial \chi_\pm^\gamma}{\partial\xi^j} \ , 
\end{equation}
where $\eta^\pm_\alpha$ are the components of the vector normal to $\Sigma$ at coordinates $\chi^\alpha_\pm$. Consequently, the second continuity condition imposed on $\Sigma$ takes the form 
\begin{equation}
  \label{16} 
     (K_{ij}^+)_\Sigma=(K_{ij}^-)_\Sigma \ .
\end{equation}

Following Nogueira and Chan \cite{Nogueira2004}, and considereing the metric given by (\ref{2}), the continuity of the first fundamental form provides
\begin{equation}
  \label{17}
    \left(\frac{dt}{d\tau}\right)_\Sigma =\frac{1}{A(r_\Sigma,t)} \ ,
\end{equation}
\begin{equation}
   \label{18}
    \left(\frac{dv}{d\tau}\right)_\Sigma^{-2}=1-2\frac{d\textbf{r}_\Sigma}{dv}-\frac{2m(v)}{\textbf{r}_\Sigma} \ ,
\end{equation}

\begin{equation}arXiv:2111.03634 [astro-ph.HE]
 \label{19}
    C(r_\Sigma,t)=\textbf{r}_\Sigma(v) = R(\tau) \ .
\end{equation}

On the other hand, the continuity of the second fundamental form leads us to obtain the mass-energy function in the form
\begin{equation}
 \label{20}
    m= \left\{\frac{C}{2}\left[1+\left(\frac{\dot{C}}{A}\right)^2-\left(\frac{C'}{B}\right)^2\right]\right\}_\Sigma \ 
\end{equation}
which represents the total energy stored inside the hypersurface. This expression is equivalent to the one given by Cahill \& McVittie \cite{Cahill}.

In addition, we can take the gravitational redshift in the form
\begin{equation}
  \label{21} 
    \left(\frac{dv}{d\tau}\right)^{-1}_\Sigma = \frac{1}{1+z_\Sigma} =  \left(\frac{C'}{B}+\frac{\dot{C}}{A}\right)_\Sigma \ ,
\end{equation}
and we can see that this expression diverges when we do
\begin{equation}
\label{22}
    \left(\frac{C'}{B}+\frac{\dot{C}}{A}\right)_\Sigma = 0\ 
\end{equation}

The continuity of the extrinsic curvature also leads us to write the identity
\begin{equation}
\label{23}
    \left(\frac{C}{2B^2}G_{11}^-\right)_\Sigma = \left(-\frac{C}{2AB}G_{01}^-\right)_\Sigma  \ .
\end{equation}
Applying the equations (\ref{10}) and (\ref{13}) to the equality above we find \cite{Nogueira2004}
\begin{equation}
\label{24}
    (P_r+4\eta\sigma-\zeta\Theta)_\Sigma = (qB)_\Sigma\ ,
\end{equation}
which represents a generalization of the junction condition $(P_r)_\Sigma=0$ in the presence of heat dissipation and viscosities.

Finally, the total luminosity of the star received by an observer at rest at infinity, denoted by $L_\infty$, is given by \cite{Lindquist}
\begin{equation}
 \label{25}
    L_\infty=-\left(\frac{dm}{dv}\right)_\Sigma=-\left[\frac{dm}{dt}\frac{dt}{d\tau}\left(\frac{dv}{d\tau}\right)^{-1}\right]_\Sigma \ .
\end{equation}

So, replacing (\ref{17}), (\ref{20}), (\ref{21}) and (\ref{24}) in (\ref{25}) we can write
\begin{equation}
 \label{26}
    L_\infty=\frac{k}{2}\left[C^2\ (P_r+4\eta\sigma-\zeta\Theta)\left(\frac{\dot{C}}{A}+\frac{C'}{B}\right)^2\right]_\Sigma \ .
\end{equation}

\end{document}